\documentstyle[12pt]{article}
\catcode`\@=11
 
\@addtoreset{equation}{section}
\def\theequation{\arabic{equation}}

\def\@normalsize{\@setsize\normalsize{15pt}\xiipt\@xiipt
\abovedisplayskip 14pt plus3pt minus3pt%
\belowdisplayskip \abovedisplayskip
\abovedisplayshortskip  \z@ plus3pt%
\belowdisplayshortskip  7pt plus3.5pt minus0pt}
\def\small{\@setsize\small{13.6pt}\xipt\@xipt
\abovedisplayskip 13pt plus3pt minus3pt%
\belowdisplayskip \abovedisplayskip
\abovedisplayshortskip  \z@ plus3pt%
\belowdisplayshortskip  7pt plus3.5pt minus0pt
\def\@listi{\parsep 4.5pt plus 2pt minus 1pt
            \itemsep \parsep
            \topsep 9pt plus 3pt minus 3pt}}
 
\def\underline#1{\relax\ifmmode\@@underline#1\else
        $\@@underline{\hbox{#1}}$\relax\fi}
\@twosidetrue
\relax
\catcode`@=12
 
\catcode`\@=11

\def\section{\@startsection{section}{1}{\z@}{3.5ex plus 1ex minus
   .2ex}{2.3ex plus .2ex}{\large\bf}}

\def\ps@headings{\def\@oddfoot{}\def\@evenfoot{}
\def\@oddhead{\hbox{}\hfill
        \makebox[.5\textwidth]{\raggedright\ignorespaces --\thepage{}--
        \hfill }}
\def\@evenhead{\@oddhead}
\def\subsectionmark##1{\markboth{##1}{}}
}
 
\ps@headings
 
\catcode`\@=12
 
\relax

\def\figcap{\section*{Figure Captions\markboth
        {FIGURECAPTIONS}{FIGURECAPTIONS}}\list
        {Fig. \arabic{enumi}:\hfill}{\settowidth\labelwidth{Fig. 999:}
        \leftmargin\labelwidth
        \advance\leftmargin\labelsep\usecounter{enumi}}}
 \relax
\def\tablecap{\section*{Table Captions\markboth
        {TABLECAPTIONS}{TABLECAPTIONS}}\list
        {Table \arabic{enumi}:\hfill}{\settowidth\labelwidth{Table 999:}
        \leftmargin\labelwidth
        \advance\leftmargin\labelsep\usecounter{enumi}}}
 \relax
\def\reflist{\section*{References\markboth
        {REFLIST}{REFLIST}}\list
        {[\arabic{enumi}]\hfill}{\settowidth\labelwidth{[999]}
        \leftmargin\labelwidth
        \advance\leftmargin\labelsep\usecounter{enumi}}}
 \relax
 
\catcode`\@=11
 
\def\marginnote#1{}
\newcount\hour
\newcount\minute
\newtoks\amorpm
\hour=\time\divide\hour by60
\minute=\time{\multiply\hour by60 \global\advance\minute by-
\hour}
\edef\standardtime{{\ifnum\hour<12 \global\amorpm={am}%
    \else\global\amorpm={pm}\advance\hour by-12 \fi
    \ifnum\hour=0 \hour=12 \fi
    \number\hour:\ifnum\minute<100\fi\number\minute\the\amorpm}}
\edef\militarytime{\number\hour:\ifnum\minute<100\fi\number\minute}
\def\draftlabel#1{{\@bsphack\if@filesw {\let\thepage\relax
  \xdef\@gtempa{\write\@auxout{\string
    \newlabel{#1}{{\@currentlabel}{\thepage}}}}}\@gtempa
    \if@nobreak \ifvmode\nobreak\fi\fi\fi\@esphack}
     \gdef\@eqnlabel{#1}}
\def\@eqnlabel{}
\def\@vacuum{}
\def\draftmarginnote#1{\marginpar{\raggedright\scriptsize\tt#1}}
\def\draft{\oddsidemargin -.5truein
        \def\@oddfoot{\sl preliminary draft \hfil
        \rm\thepage\hfil\sl\today\quad\militarytime}
        \let\@evenfoot\@oddfoot \overfullrule 3pt
        \let\label=\draftlabel
        \let\marginnote=\draftmarginnote
   
\def\@eqnnum{(\theequation)\rlap{\kern\marginparsep\tt\@eqnlabel}%
\global\let\@eqnlabel\@vacuum}  }
\def\preprint{\twocolumn\sloppy\flushbottom\parindent 1em
        \leftmargini 2em\leftmarginv .5em\leftmarginvi .5em
        \oddsidemargin -.5in    \evensidemargin -.5in
        \columnsep 15mm \footheight 0pt
        \textwidth 250mmin      \topmargin  -.4in
        \headheight 12pt \topskip .4in
        \textheight 175mm
        \footskip 0pt
\def\@oddhead{\thepage\hfil\addtocounter{page}{1}\thepage}
        \let\@evenhead\@oddhead \def\@oddfoot{} \def\@evenfoot{} 
}
\def\titlepage{\@restonecolfalse\if@twocolumn\@restonecoltrue\onecolumn
     \else \newpage \fi \thispagestyle{empty}\c@page\z@
        \def\thefootnote{\fnsymbol{footnote}} }
\def\endtitlepage{\if@restonecol\twocolumn \else  \fi
        \def\thefootnote{\arabic{footnote}}
        \setcounter{footnote}{0}}  
\catcode`@=12
\relax
 
\def\ps@headings{\def\@oddfoot{}\def\@evenfoot{}
\def\@oddhead{\hbox{}\hfill
        \makebox[.5\textwidth]{\raggedright\ignorespaces --\thepage{}--
        \hfill }}
\def\@evenhead{\@oddhead}
\def\subsectionmark##1{\markboth{##1}{}}
}
 
\ps@headings
 
\relax

\def\firstpage#1#2#3#4#5#6{
\begin{document}
\def\beq{\begin{equation}} 
\def\eeq{\end{equation}} 
\def\bea{\begin{eqnarray}} 
\def\eea{\end{eqnarray}} 
\def\bq{\begin{quote}} 
\def\eq{\end{quote}}
\def\ra{\rightarrow} 
\def\lra{\leftrightarrow} 
\def\ups{\upsilon}
\def\bq{\begin{quote}} 
\def\eq{\end{quote}}
\def\ra{\rightarrow} 
\def\un{\underline}
\def\ov{\overline}
\newcommand{\cm}{Commun.\ Math.\ Phys.~}
\newcommand{\prl}{Phys.\ Rev.\ Lett.~}
\newcommand{\pr}{Phys.\ Rev.\ D~}
\newcommand{\pl}{Phys.\ Lett.\ B~}
\newcommand{\ibar}{\bar{\imath}}
\newcommand{\jbar}{\bar{\jmath}}
\newcommand{\np}{Nucl.\ Phys.\ B~}
\newcommand{\F}{{\cal F}}
\renewcommand{\L}{{\cal L}}
\newcommand{\A}{{\cal A}}
\def\154{\frac{15}{4}}
\def\153{\frac{15}{3}}
\def\32{\frac{3}{2}}
\def\254{\frac{25}{4}}
\begin{titlepage}
\nopagebreak
\title{\begin{flushright}
        \vspace*{-1.8in}
        {\normalsize IOA-TH/97-004}\\[-9mm]
        {\normalsize hep-ph/9703338}\\[4mm]
\end{flushright}
\vfill
{#3}}
\author{\large #4 \\[1.0cm] #5}
\maketitle
\vskip -7mm     
\nopagebreak 
\begin{abstract}
{\noindent #6}
\end{abstract}
\vfill
\begin{flushleft}
\rule{16.1cm}{0.2mm}\\[-3mm]
$^{\star}${\small Research supported in part by the EEC under the 
\vspace{-4mm} TMR contract ERBFMRX-CT96-0090 and by ${\Pi}$ENE${\Delta}$
91E${\Delta}$300.}\\ 
March 1997
\end{flushleft}
\thispagestyle{empty}
\end{titlepage}}
 
\def\simlt{\stackrel{<}{{}_\sim}}
\def\simgt{\stackrel{>}{{}_\sim}}
\date{}
\firstpage{3118}{IC/95/34}
{\large\bf  Constraints on Leptoquark Masses and Couplings
from Rare Processes and Unification  $^{\star}$} 
{ G. K. Leontaris and J. D. Vergados }
{\normalsize\sl
Theoretical Physics Division, Ioannina University,
GR-45110 Ioannina, Greece.}
{Motivated by the recent experimental H1 and ZEUS data at HERA,
which have reported evidence for leptoqark production
at $\sqrt{s}={314}$ GeV with a mass at $m_{D}=200$GeV 
we consider its implications in unified supersymmetric
theories. We also present calculations for leptoquark 
production incorporating the existing limits from other exotic 
reactions on its couplings and other relevant parameters.}

\newpage
Recently, H1\cite{hera} and ZEUS\cite{zeus}  experiments have
reported an excess of $e^+p$ deep inelastic scattering events
at very high $Q^2$ and large $x$. In the past, there have appeared many 
suggestions which may interprete the high $Q^2$ - events.
{}For example, the reported excess  could be  explained  with the existence
of  R - parity violating interactions\cite{rp}  in supersymmetry.
These data, could also be compatible with a narrow state suggestive
of a new particle possessing both lepton and baryon  quantum numbers 
( leptoquark)  with a mass $m_{D} \sim 200GeV$. This exciting 
possibility, at the time of writing this work,  resulted in a
 number of very interesting theoretical
considerations\cite{aeglm,new}. In the present work, we will attempt
to describe the properties of such a state, in terms of
isosinglet colour triplet superfields $D,D^c$ by incorporating
known limits from other exotic processes.

It is interesting that various extentions of the Minimal Supersymmetric
Standard Model ( MSSM)  predict the existence of several kinds  of
new particles in addition to the known quarks and leptons.
Among them, the leptoquarks as well could be proposed as a 
signal for new physics beyond MSSM.
The intoduction of any new particle in the minimal
theory has important implications which should not be overlooked.
Two main concerns should be the following. New particles create
new interactions in the theory which often lead to severe constraints
on their masses and Yukawa couplings. Second, we know that the great
and impressive success of the 
supersymmetric theories is intimately related to the unification of
 the fundamental forces at a large energy scale.
Thus, the appearance of relatively light (${\cal O}$(TeV) ) 
states in the supersymmetric spectrum will have a significant
effect on the gauge coupling running and therefore they might
spoil the unification scenario. 

In this note,
we would like to address these two points in the context of
unified theories whose low energy limit is a supersymmetric
model which in addition to the MSSM spectrum has also  
a pair of leptoquarks $D,D^c$ sitting in $(3,1,-\frac 13)$ and 
$(\bar{3},1,\frac 13)$ representations of the standard model
gauge symmetry.
We will see that due to the possible existence of Yukawa couplings
of the particles $D,D^c$ with the ordinary matter, rare processes
put rather stringent constraints on the related couplings. On the
other hand, if we wish to retain the successful unification scenario,
we find that we are forced to include in the spectrum new particles
which become massive at some intermediate scale.

The basic superpotential couplings which result to the fermion
masses in the MSSM are the following
\bea
{\cal W} &=& \lambda_1 Q u^c h_1 +\lambda_2 Q d^c h_2 +
              \lambda_3 L e^c h_2 +\lambda_4 \phi_0 h_1h_2
\label{w}
\eea
where $Q, u^c, d^c , L,  e^c$ are the quark and lepton
superfields and $h_{1,2}$ the standard higgses. $\phi_0$ is
a singlet which realises the higgs mixing.
Now, we may assume in
addition the existence of $D, D^c$ particles.
There are two types of couplings which can exist in the superpotential.
These are,
\bea
 {\cal W}_1 &=& \lambda_5 Q Q D +\lambda_6 u^c d^c D^c 
\label{w1}
\eea
and 
\bea
{\cal W}_2 &=& \lambda_7 D^c Q L + \lambda_8 D u^c e^c +
           \lambda_{8}' D d^c \nu^c\label{w2}
\eea
where we have assumed that $\nu^c$ is the right handed neutrino.
If all terms of (\ref{w1},\ref{w2}) are present in the 
superpotential, the related Yukawa couplings should be unnaturally
small in order to prevent fast proton decay\cite{je,lv0}.
With a suitable discrete symmetry\cite{je} it is possible 
to prevent one of ${\cal W}_1,{\cal W}_2$. It is clear therefore,
that  the experimental findings at H1 and ZEUS,
provided that they are not swept out in future runs, 
suggest that we  retain ${\cal W}_2$. In this case $D, D^c$ carry
both lepton and baryon quantum numbers (leptoquarks). 

It is known that the particle content of the MSSM  allows the
three gauge couplings to attain a common value at a high scale,
of the order $M_U\sim 10^{16}GeV$.  The introduction of
massless states beyond those of the minimal spectrum change drastically
the evolution of the gauge couplings. Thus, if we assume the
existence of a pair of leptoquarks remaining massless down to 
the weak scale, in order that the idea of unification remains intact at
some high scale (not necessarily the same as $M_U$), additional
contibutions to the beta functions are needed to compensate  
for the leptoquark pair and yield a correct prediction for
the weak mixing angle. In order to
clarify this point let us assume that unification takes place at the scale
$M_U$ and assume in addition the existence of extra matter fields.
Various kinds of exotic fields are present in unified theories,
in particular in superstring derived models.  In the
present work however, we will assume only additional multiplets
which carry quantum numbers like those of ordinary quarks and leptons.
Thus, let us denote with $n_{D}$ the number of the leptoquarks 
and $n_{Q},n_{u^c}, n_{\ell}, n_{e^c}$ those of possible 
 additional left and right handed type quarks,
 left leptons and right handed electrons respectively. {} For
our purposes it is enough to consider only $n_{Q}, n_{e^c}, n_{D}$.
Now, having in mind a unified model like flipped $SU(5)$ etc,
 the low energy measured quantities in terms of the extra matter 
fields and the scale $M_U$ at the one - loop level, are\cite{aekn}
\bea
 \log\frac{M_U}{M_Z}&=&\frac{\pi}{10\alpha}
(1-\frac{8}{3}\frac{\alpha}{\alpha_3}) +\frac{1}{20\pi}
(\vec{n}_{Q}+\vec{n}_D-\vec{n}_{e^c})
\cdot \vec{\cal L} \label{mu}\\
\sin^2\theta_W &=& \frac 15 + \frac{7}{15}\frac{\alpha}{\alpha_3}
+\frac{\alpha}{20\pi} (7\vec{n}_{Q}-3\vec{n}_D-2\vec{n}_{e^c})
\cdot \vec{\cal L}\label{sw2}
\eea
where  $\vec{n}_{r}\cdot\vec{\cal L} =
\sum_I n^I_r \log(M_I/M_{I-1})$ takes into
account the number of multiplets which remain massless at various
possible intermediate scales $M_I$. It is clear from the 
formulae (\ref{mu},\ref{sw2}) that if a leptoquark pair 
remains in the massless spectum this will alter both the unification
scale and $\sin^2\theta_W$. Thus, for example, assuming $n_D=2$,
with no other extra multiplets ($ n_{Q}= n_{e^c}=0$),
from Eq.(\ref{mu}) we find that $M_U\sim 7\times 10^{17}$GeV.
This is welcome as it is of the order of the string scale. However,
in this simple case,  the weak mixing angle form
Eq.(\ref{sw2}) does not have the right value
($\sin^2\theta_W(m_W)\sim 0.21 !$). Thus, it is clear, that additional
fields must coexist with the leptoquarks in order to cancel
their contribution into the above equations and allow
unification consistent with the correct value of $\sin^2\theta_W$.
To pursue our argument further, let us consider two simple
cases:\\
{\it  i)}The unification scale is the  same as in the MSSM case,
while the possible additional states remain also down to
the elecroweak scale, or\\
{\it  ii) } we assume that the additional matter fields
  may receive a mass at some intermediate scale. We will see
in this second case that a natural scenario implies a unification
mass of the order of the String scale.

 Let us start with the analysis of the  case {\it  i)}.
 Combining the above equations
we may eliminate the scale $M_U$ and express the deviation  of
the weak mixing angle from its MSSM value as follows
\beq
\delta \sin^2\theta_W = 
    \frac{7{n}_{Q}-3{n}_D-2{n}_{e^c}}{ 
           20-n_{Q}-n_D+n_{e^c}}
    \frac{1}{10}\left(1-\frac 83\frac{\alpha}{\alpha_3}\right)
    \eeq
{}From  the fact that $n_i$ are integers, 
it can be seen that $\delta \sin^2\theta_W $
can be within the accepted bounds only if the quantity 
$7{n}_{Q}-3{n}_D-2{n}_{e^c}$ is zero. For $n_D=2$, this
happens if $n_{Q}=2 , n_{e^c}=4$. From Eq.(\ref{mu}) we also
note that the scale $M_U$ is the same as in the minimal
case, 
\beq
M_U = M_Z Exp\left(\frac{\pi}{10\alpha}
\left\{1-\frac{8}{3}\frac{\alpha}{\alpha_3}\right\} \right)
\sim 10^{16} GeV.
\eeq
It is clear that the only contribution to the beta fuctions
which has the potential to cancel the $D,D^c$ effect comes
from the vector like quark superfields $Q' +\bar{Q}'$
\cite{aekn} as these are the only quantities entering with
the opposite sign.

{\it ii)} In the second possibility, we may consider that the 
contribution of the leptoquarks in the beta function coefficients is 
compensated by new states which become massive at an intermediate
scale. This scenario may be realised consistently if we assume
for example that an equal number of left type quarks $Q'$ and
right handed electrons ${e^c}'$ , i.e. $n_{e^c}=n_Q$,
become massive at some intermediate scale $M_I$. From Eq.(\ref{sw2})
we find that the mass scale $M_I$ is
\beq
M_I = \left(\frac{m_D}{M_U}\right)^{\frac{3}{5}\frac{n_D}{n_Q}}
M_U
\label{mi}
\eeq
{}For $n_Q =2$, this gives $M_I\sim 10^8$GeV. Table 1 shows
 the unification scale and the mass of the new states for
two representative cases.
\begin{table}
\centering
\begin{tabular}{|l|c|r|c|c|} \hline
$\alpha_3$& $\sin^2\theta_W$ & $M_U$/GeV      &  $M_I$/GeV&$ m_{D}$/GeV
 \\ \hline
0.110      &    0.2330         & $4.9\cdot 10^{17}$&  
 $2.8\cdot 10^8$ & 200\\
0.115      &    0.2317         & $7.0\cdot 10^{17}$&
$ 3.3\cdot 10^8$& 200  \\
\hline
\end{tabular}
\caption{\it The Unification scale and the mass of the Vector Quark 
multiplet for two values of
$\alpha_3$,  and a leptoquark mass $m_D$ = 200 GeV.}
\label{table:I}
\end{table}

 The terms in the superpotential of eq.(\ref{w2}) lead to some very
interesting phenomenology a major part of which has been explored in a
previous paper
\cite{lv0}.
Here we will review the essential conclusions of that paper and extend 
it to cover recent developments.  Thus, 
 the first term of eq. (\ref{w2}) leads to quark lepton fusion into
a leptoquark which for left handed fermions takes the form:
\beq
\L_{eff}=\lambda_7D^{c}{\bar e{c}_{R}}u_L + H.C.
\label{eq:2.1}
\eeq
 while for right handed fermions one obtains:
\beq
\L_{eff}=\lambda_8D{\bar e{c}_{L}}u_R + H.C.
\label{eq:2.2}
\eeq
Both of them lead to $e^{+}{\bar u}$ and $e^{-}u$ fusion to 
a leptoquark. The first can occur  via the sea antiquarks
in the proton as we shall see below and is expected to
proceed at a smaller rate compared to the second. For a proton
target the electron beam is favored. Thus
the leptoquarks generated from sea-antiquarks allow
larger couplings than those generated in collisions with
valence quarks. 

 Since the mass of the leptoquark has been constrained from the recent
experiments we will attempt to constrain the flavor diagonal couplings
$\lambda_7$ and $\lambda_8$. The ordinary
$\beta$- decay leads to the bound \cite{lv0}
\beq
\lambda_7^2 \le \sqrt{2}G_{F}m^2_{D^c}\times 10^{-2}.
\label{eq:2.3}
\eeq
which for $m_{D^c}=200 GeV$ yields $\lambda_7 \leq 2.6 
\times 10^{-2}$.
 The parameter $\lambda_8$ cannot be similarly constrained since
now $\beta$-decay is the combined effort of $\lambda_8,
\lambda_8'$. From the
non-observation of $\beta$ - decay involving right - handed currents
one can set the following limit
\beq
\lambda_8\lambda_8'\le \sqrt{2}G_{F}m^2_{D} \times 10^{-2}.
\label{eq:2.4}
\eeq
 If, however, we make the reasonable assumption that $\lambda_8=
\lambda^{\prime}_8$ we obtain the same
 limit as above, namely $\lambda_8 \leq 2.6\times 10^{-2}$.\
Additional constaints can be obtained from double beta decay once there is 
a coupling between $(\tilde D,\tilde d)$ and/or $({\tilde D^c},\tilde d^c)$.
This can arise out of the couplings $\lambda_7$ and $\lambda^{\prime}_8$
 once the
s-neutrinos acquire vacuum expectation values. Contrary to our previous
work we will not assume here  the mass of $\tilde D$ to be much 
larger than the mass of ${\tilde d}$. Thus the lepton violating parameter,
since the gluino mediated process was found dominant\cite{lv0}, 
takes the form
\beq
\eta_{{\tilde D},{\tilde g}} =\frac{\pi \alpha_{s} m_{p}}{3m_{{\tilde g}}}
 \left(\frac{\lambda_7 s_{L}c _{L}(m^2_{2L}-m^2_{1L})
}{G_{F}m^2_{1L}m^2_{2L}}\right)^2 
\label{eq:2.5}
\eeq
\beq
\eta_{{\tilde D^c},{\tilde g}} =
\frac{\pi \alpha_{s} m_{p}}{3m_{{\tilde g}}}
 \left(\frac{\lambda_7 s_{R}c _{R}(m^2_{2R}-m^2_{1R})
}{G_{F}m^2_{1R}m^2_{2R}}\right)^2
\label{eq:2.6}
\eeq
where $m_{1L},m_{2L}$ are the ${\tilde D},{\tilde d}$
 eigenstates and $m_{1R},m_{2R}$ are the 
${\tilde D^c},{\tilde d^c}$ eigenstates, $m_{\tilde g}$ 
is the gluino mass and
\bea
s_{L,R} = \sin\theta_{L,R}\, , & c_{L,R} = \cos\theta_{L,R}
\eea
where $\theta_{L,R}$ is the corresponding mixing angle.  Since 
the effective four quark interaction is not of the $V-A$ type
but of the $S,P,T$ variety, there are ambiguities in going from
the quark to the nucleon level\cite{JDV}. Therefore, large 
 ambiguities are expected in extracting limits from the non-
observation of $\beta\beta_{0\nu}$- decay. In the spirit
of \cite{JDV,lv0,klap}, using  
  the most recent experimental data on $^{76}Ge$\cite{JDV}
we obtain 
\beq
\eta_{{\tilde D^c},{\tilde g}} \leq 4\times 10^{-7} 
\label{eq:2.7}
\eeq
 with a similar limit on $\eta_{{\tilde D^c},{\tilde g}}$.
 The limit (\ref{eq:2.7}) can be converted to a limit on
 the Yukawa coupling once the scalar quark and gluino mass
 eigenstates are known. Using renormalisation group analysis,
  for $\tan\beta\sim 10$  we find ${m}_{\tilde d}\sim 276$GeV
 and ${m}_{\tilde g}\sim 272$ GeV which turns to a limit
 $\lambda_7\le 0.3 - 0.06$, for  mixings $0.1 - 0.7$.
 A  similar bound is obtained for other consistent
 choices of the scalar mass spectrum in the allowed 
 parameter space.
 It turns out that $\beta\beta_{0\nu}$-decay in this
 case does not improve the ordinary $\beta$-decay bound obtained
 previously.

 We come now to the calculation of the cross section\cite{bw} in
 the case of a theory with a leptoquark coupling. The
 cross section of a leptoquark production in the 
spin zero case is equal to
\bea
\sigma&=&\frac{\pi}{2s}\frac{\lambda^2}2f\left(\frac{m^2}{{s}}\right)
\nonumber\\
       &=& 0.31\, {\rm pb}\, 
        \left(\frac{\lambda}{0.01}\right)^2
        f\left(\frac{m^2}{{s}}\right)
\label{cs}
\eea
Using the quark distributions of ref.\cite{grv}, one obtains the
fusion cross sections as a function of $m_D$,  $s$ and
the leptoquark coupling $\lambda$. Representative computations for
$\lambda$ values respecting the $\beta$ and $\beta\beta$ - decay
limits are shown in table 2.
\begin{table}
\centering
\begin{tabular}{|l|c|c|c|c|} \hline
$$& ${ m_D :}$ & 100GeV    &  200GeV & 250 GeV
 \\ \hline
$\lambda=0.01$& $\sigma(e^+\bar{u})$: & 0.27 & $1.6\times 10^{-3}$
                                            & $3.3\times 10^{-5}$
 \\ $\lambda=0.02$&                 & $1.10$ & $6.4\times 10^{-2}$
                                            & $1.3\times 10^{-4}$
 \\ \hline
$\lambda=0.01$& $\sigma(e^-    {u}):$ & 1.78 & $1.6\times 10^{-1}$
                                            & $1.8\times 10^{-2}$
 \\ $\lambda=0.02$ &               & $7.20$  & $6.4\times 10^{-1}$
                                            & $7.2\times 10^{-2}$
 \\
\hline
\end{tabular}
\caption{\it $\sigma(e^+\bar{u})$ and  $\sigma(e^-    {u})$
for three cases of $m_D$ = 100, 200 and 250 GeV, for
the indicated values of the Yukawa coupling $\lambda $
($\lambda =\lambda_7$ or $\lambda = \lambda_8$).}
\label{table:II}
\end{table}

In conclusion, we have shown that the introduction of the
leptoquark superfields $D,D^c$ and its superpotential
couplings of the term ${\cal W}_2$ in Eq.(\ref{w2}) introduced
to explain the experimental data, does not lead to any inconsistencies
with the unification of gauge couplings provided there are additional 
left quark vector-like multiplets which become massive at some 
intermediate scale. Furthermore, utilizing the existing constraints
on the leptoquark couplings from exotic reactions, recent experimental
findings of H1 and ZEUS can be adequately explained.

Note Added: As this work was being prepared, we have also noticed
several new papers dealing with the interpretation of the HERA
events\cite{new1}

\end{document}